\documentclass{aa}
\let\linenumbers\relax
\let\nolinenumbers\relax

\usepackage{graphicx}
\usepackage{txfonts}
\usepackage{amsmath}
\usepackage{amssymb}
\usepackage{url}

\graphicspath{{images/}{./}}

\begin{document}

\title{\textsc{GTLS}: A GPU-accelerated method for periodic transit detection}

\author{
Quanquan Hu\inst{1,2}
\and Jian Ge\inst{1}\corrauth{jge@shao.ac.cn}
\and Luoxi Jin\inst{1,2}
\and Kevin Willis\inst{3}
}

\institute{
Shanghai Astronomical Observatory, Chinese Academy of Sciences, 80 Nandan Road, Shanghai 200030, China
\and University of Chinese Academy of Sciences, 19A Yuquanlu, Beijing 100049, China
\and Science Talent Training Center, Gainesville, FL 32606, USA
}

\date{}

\abstract
{Computational efficiency is a critical requirement for transit searches in modern large-scale photometric surveys.}
{We present Graphics Processing Units (GPU) Transit Least Squares (\textsc{GTLS}), a GPU-accelerated implementation of the Transit Least Squares algorithm designed to reduce the computational cost of periodic transit detection while preserving TLS-like sensitivity to transit-shaped signals.}
{\textsc{GTLS} parallelizes the dominant steps of the TLS search, including phase folding, transit-duration evaluation, moving-window depth estimation, and $\chi^2$ calculation. Using \textit{Kepler-like} long-cadence light curves and synthetic \textit{Kepler-like} time series, we benchmark \textsc{GTLS} against the reference CPU implementation of TLS and the GPU-based BLS implementation in \texttt{cuvarbase}.}
{On an AMD Ryzen 9 7950X CPU and an NVIDIA RTX 4090 GPU, \textsc{GTLS} processes a 3000-day synthetic light curve in approximately 138 seconds, compared with 3289 seconds for TLS. With two RTX 4090 GPUs, the runtime is reduced to approximately 79 seconds. In recovery tests, \textsc{GTLS} achieves detection performance statistically consistent with TLS, with a precision of 9.3 percent and recall of 79.4 percent, compared with 9.4 percent and 81.1 percent for TLS.}
{These results demonstrate that \textsc{GTLS} enables efficient TLS-style searches for large photometric data sets from \textit{Kepler}, \textit{TESS}, \textit{PLATO}, ET, and future missions. The source code is publicly available.}

\keywords{methods: data analysis -- planets and satellites: detection}

\maketitle



\section{Introduction}
The discovery of exoplanets represents one of the most significant advances in modern astronomy. A large fraction of known exoplanets has been discovered using the transit method, as summarized by the NASA Exoplanet Archive \citep{akeson2013}. With the increasing number of space-based search programs such as \textit{Kepler} \citep{Borucki_2010}, K2 \citep{Howell_2014}, TESS \citep{Ricker_2014}, the upcoming PLATO \citep{Rauer_2014, Rauer_2025}, and ET \citep{Ge_2022a, Ge_2022b, Ge_2022c, Ge_2024a, Ge_2024b}, a large number of light curves have been or will be produced. Efficient processing of light curve data and data analysis \citep[e.g.,][]{Wang_2024a} has become an increasingly urgent need.

The Box Least Squares (BLS) algorithm \citep{Kov_cs_2002} has been widely used as a standard method to detect transit signals in large light curve data sets. The BLS algorithm approximates the transit as a box-shaped signal with a uniform in-transit depth, neglecting stellar limb darkening and planetary ingress/egress. This simplification reduces the computational difficulty of finding transit signals and provides a consistent basis for evaluating targets with different ingress and egress times. However, due to this simplification, the detection efficiency of the BLS for Earth-sized planets is low because the depth of the transiting signal may be comparable to instrumental or stellar noise (\cite{Hippke_2019a}, hereafter H19). To address this limitation, the Transit Least Squares (TLS) algorithm opts for a transit-like search function rather than a box function. The TLS algorithm chooses a representative transit model from confirmed \textit{Kepler} exoplanets to construct the search function (H19). Therefore, TLS is more efficient than BLS in searching for transit signals from Earth-sized planets. This was also confirmed by the PLATO group in a comparison of performance from different pipelines \citep{Canocchi_2023}.

The classical Box-fitting Least Squares (BLS) algorithm and many widely used implementations are CPU-based, which limits the speed-ups achievable through parallelism. The \texttt{cuvarbase} package proposed by \citet{Hoffman_2017} provides a GPU-accelerated implementation of BLS and delivers substantial runtime improvements relative to typical CPU implementations. \citet{Kunimoto_2023} reports that this GPU implementation yields detection results comparable to those obtained with standard BLS. We therefore adopt \texttt{cuvarbase} as our BLS baseline for the performance comparisons presented in this paper.

Due to the application of the transit-like signal search function, the TLS algorithm is more complex than BLS. Although H19 has parallelized the TLS algorithm on a CPU platform, it is too slow to be used in large-scale data processing and analysis \citep{Wang_2024a}. For a typical \textit{Kepler} long cadence (LC) light curve, TLS takes about $10^3$ seconds (or 17 minutes) to complete the search in a \textit{Kepler} light curve on an Intel i7-7700k CPU (H19). Considering \textit{Kepler} has collected light curves from about 200,000 targets, it will take more than 2 years on a single desktop computer to complete the search, which is not practical for large-scale data processing, transit signal searching, and signal analysis.

In this work, we present a GPU-accelerated implementation of TLS for speeding up periodic transit searches using a GPU-accelerated TLS called the GPU transit least squares (\textsc{GTLS}) algorithm. Running on modern GPUs, \textsc{GTLS} provides an order-of-magnitude speed-up relative to the reference CPU implementation of TLS for periodic transit searches.

We implement the host-side components in Python and use CuPy \citep{cupy_learningsys2017} to execute the computationally intensive kernels on NVIDIA GPUs via CUDA. In particular, the GPU accelerates the dominant cost of the algorithm: evaluating the transit model over the period--duration grid.

The key contribution of this work is not simply the use of the GPU, but the adaptation of the TLS search architecture---including period/duration grid construction, phase folding, padding, moving-window depth estimation, and $\chi^2$ evaluation---to a highly parallel GPU framework. This enables TLS-like transit-model searches to be performed at speeds approaching or exceeding GPU-accelerated BLS implementations, while preserving the improved transit-shape matching of TLS.

The organization of this paper is as follows: Sect.~2 introduces the methods and details. Sect.~3 describes performance, including data preparation, preprocessing, accuracy, speed, and their comparison with other methods. In Sect.~4, we present a discussion of our method, followed by conclusions in Sect.~5. 

\section{Methods}

\subsection{Overview of the \textsc{GTLS} algorithm}
\label{sec:overview_of_the_gtls_algorithm}

\textsc{GTLS} is a transit-detection method based on the TLS method, which significantly improves processing speed by utilizing the parallel computing power of GPUs. The TLS method iterates through various combinations of period, duration, and $T_0$ to calculate the chi-squared ($\chi^2$) value, which is used to identify transit events. Traditional TLS algorithms are computationally intensive and time-consuming, whereas \textsc{GTLS} accelerates these steps with GPUs, enabling simultaneous processing of multiple combinations of period, duration, and $T_0$, greatly enhancing detection efficiency. Particularly, in steps such as light curve folding, padding, moving average, and $\chi^2$ calculation, parallelization on GPUs significantly speeds up the process, allowing large amounts of data to be processed quickly and optimizing the identification of transit signals.

Similar to existing transit detection methods like TLS and BLS, \textsc{GTLS} requires determining the period, duration, and $T_0$ to accurately locate the transit event. As a least squares method, \textsc{GTLS} iterates through combinations of period, duration, and $T_0$ to identify the most probable transit and provides confidence metrics such as signal detection efficiency (SDE) and signal-to-noise ratio (SNR). Given a light curve and a period, the light curve can be folded according to the period. Given a duration, the ideal transit signal can be generated by calculating the number of data points in the transit signal and the flux of each point. The ideal transit signal is then scaled to match the average flux of the target curve. Therefore, the first task in the \textsc{GTLS} algorithm is to determine the grids to be searched for period and duration. The grid is a combination of period and duration, and the determination of the grid for iteration is directly related to the accuracy and speed of the search. We will discuss in detail the grids for period and duration determined by \textsc{GTLS}, as well as the search method for $T_0$ in Sects.~\ref{sec:gtls_parameter_grid} and~\ref{sec:search_point_by_jumping_point_and_secondary_search}.

In \textsc{GTLS}, the orbital period, transit duration, and transit epoch (T$_0$) are the key search parameters that determine how the light curve is phase-folded and how individual transit events are aligned and stacked. The orbital period controls the separation between successive transits. Because small period errors accumulate more rapidly for short period signals, denser period sampling is required at short periods. In contrast, long-period signals are less sensitive to small period errors over a fixed observational baseline, allowing a coarser period grid at longer periods.

The transit duration defines the expected width of each transit event. Since transit duration is statistically correlated with orbital period, with shorter-period planets generally producing shorter-duration transits and longer-period planets producing longer-duration transits, the duration grid should be constructed using the expected period--duration relation for transiting systems. This improves the match between the search template and the true transit morphology.

The transit epoch (or the mid-transit time, T$_0$), which specifies the reference time of a transit event, is critical for aligning the folded signal. \textsc{GTLS} treats T$_0$ as an independent search parameter. Candidate T$_0$ values are first identified through a coarse grid search and then refined locally to improve alignment precision. This hierarchical search strategy improves both the accuracy and computational efficiency of transit detection.

\begin{figure*}
\includegraphics[width=\textwidth]{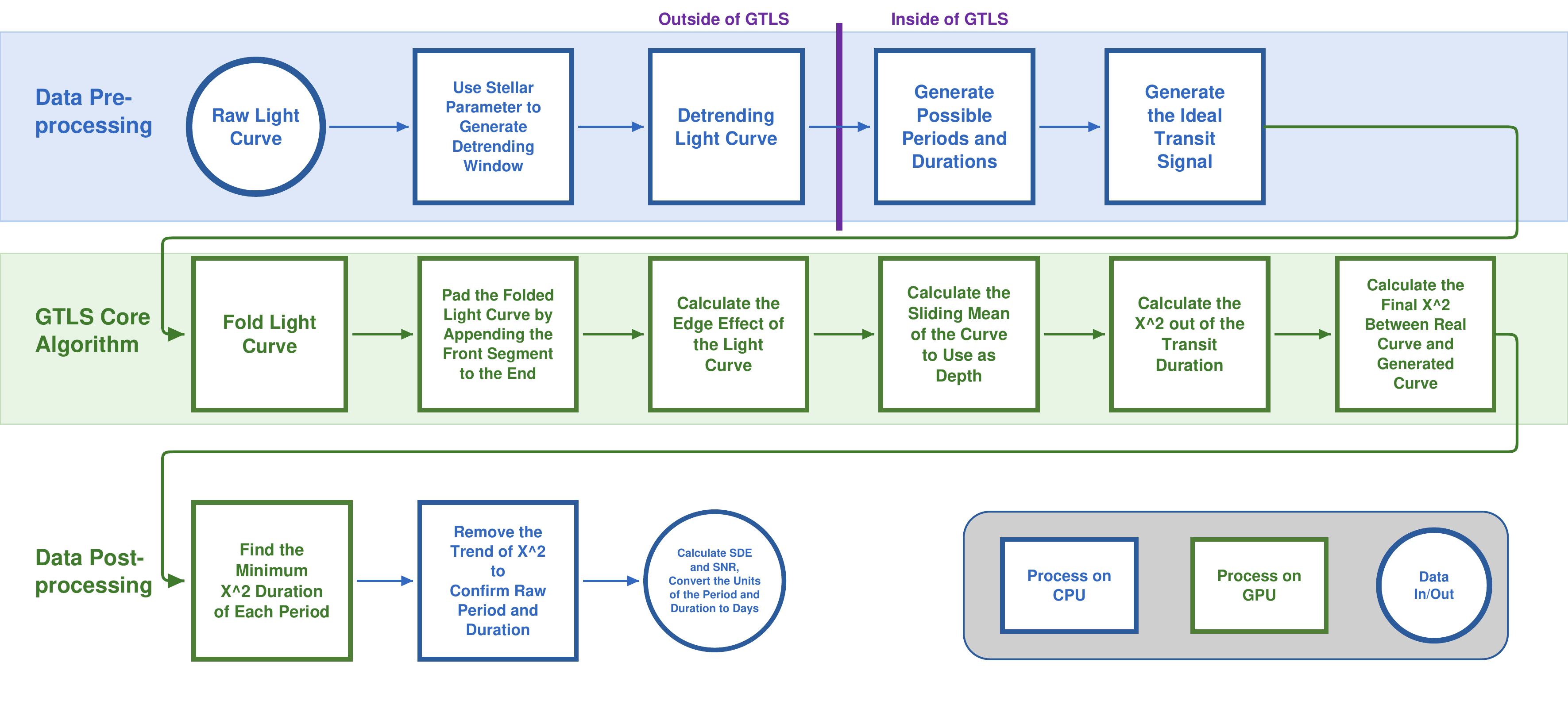}
\caption{Overview of the processing pipeline from the raw light curve to the \textsc{GTLS} output. Part of the preprocessing is performed outside the \textsc{GTLS} framework.}
\label{fig:gtls_structure_figure}
\end{figure*}

With appropriately designed grids for period, duration, and T$_0$, \textsc{GTLS} reduces computational costs while preserving accurate phase folding and signal stacking for each trial transit model. The overall architecture of \textsc{GTLS} is summarized in Fig.~\ref{fig:gtls_structure_figure}.

To identify the most probable transit event, \textsc{GTLS} uses the chi-squared ($\chi^2$) value as the basic evaluation metric. Both SDE and SNR are derived based on $\chi^2$. Specifically, the $\chi^2$ value is calculated by comparing the ideal transit signal with the real light curve using the classical least squares method \citep{Norman_1966}. Sect.~\ref{sec:calc_chi2} will discuss the details of the $\chi^2$ calculation.

Besides the calculation of $\chi^2$, the core algorithm of \textsc{GTLS} also includes light curve folding, light curve padding, and moving average computation. These three steps are all preparatory for the $\chi^2$ calculation. Sect.~\ref{sec:highly_parallelized_implementation_on_gpu} will discuss the implementation details of these steps on the GPU.

In a typical \textsc{GTLS} usage scenario, data preprocessing, such as detrending, is also required. However, since this part involves various processing methods and is not the main focus of \textsc{GTLS}, it is not included in the code provided by \textsc{GTLS}. The preprocessing method used in this paper is designed to be simple and computationally efficient and is described in Sect.~\ref{sec:sec_performance_accuracy_KOI}.

\subsection{\textsc{GTLS} parameter grid}
\label{sec:gtls_parameter_grid}

\textsc{GTLS} searches for transit signals by folding the real light curve according to different possible periods and then compares the folded light curve with the ideal transit signal. The ideal transit signal is generated based on a given duration.
To reflect the real astrophysical situation and reduce computational complexity, we use specific parameter grids to limit the search range of the period, duration, and $T_0$.

\subsubsection{Period sampling}
\label{sec:period_sampling}

Period and duration are core parameters that describe the fundamental characteristics of transit signals, determining the phase folding of light curves and the stacking of signals. The period refers to the time interval between consecutive transit events.

\cite{Ofir_2014} provides an example of the period grid for the BLS algorithm. The period grid for the TLS algorithm is similar, and we will follow this procedure.

To detect periodic phenomena in the light curve, the maximum detectable period $P_{max}$ is $S/2$, where $S$ is the length of the light curve, and the minimum frequency $f_{min}$ is
\begin{equation}
	f_{min} = \frac{1}{P_{max}} = \frac{2}{S}.
\end{equation}

Keplerian dynamics give the relationship between orbital period $P$ and orbital semi-major axis $a$:
\begin{equation}
	P^2 = \frac{4\pi^2}{GM}a^3.
\end{equation}

Considering the minimum orbital semi-major axis $a_{min}$ to be the Roche limit, assumed to be $3R$, where $R$ is the radius of the star \citep{Guillochon_2011}. Therefore, the maximum frequency $f_{max}$ is
\begin{equation}
	\label{eq:period_sampling_fmax}
	f_{max} = \frac{1}{2\pi}\sqrt{\frac{GM}{(3R)^3}},
\end{equation}
where $M$ is the mass of the star, and $G$ is the gravitational constant.

Unlike uniform sampling of trial frequencies, \cite{Ofir_2014} derived a cubic sampling of the trial frequencies, which is considered optimal. The sampling frequency is
\begin{equation}
	f_i = \left(\frac{A}{3}i + C\right)^3,
\end{equation}

where $f_i$ is the $i$-th sampling frequency, $i=1,\dots,N$, $N$ is the total number of samples, and $C$ is a constant that ensures the minimum frequency is $f_{min}$. The value of $C$ is
\begin{equation}
	\label{eq:period_sampling_C_A}
	\begin{array}{l}
	C = f_{min}^{1/3} - \frac{A}{3} \\
	A = \frac{(2\pi)^{2/3}}{\pi}\frac{R}{(GM)^{1/3}}\frac{1}{S\cdot OS},
	\end{array}
\end{equation}
where $S$ is the oversampling factor.

Considering the maximum frequency $f_{max}$, the number of sampling frequencies is:
\begin{equation}
	\label{eq:period_sampling_N}
	N = \left(f_{max}^{1/3} - f_{min}^{1/3} + \frac{A}{3}\right) \frac{3}{A}.
\end{equation}

In the period grid, users can input three parameters: stellar mass $M$, stellar radius $R$, and the oversampling parameter $S$. Users can adjust the upper limit of the search frequency and the search density through different inputs. In \textsc{GTLS}, the default values for these three parameters are the same as those used in TLS: the stellar mass and radius are set to be identical to the Sun, and the oversampling parameter is set to 3.

By substituting the default parameters of \textsc{GTLS} into Eq.~(\ref{eq:period_sampling_fmax}), we obtain $f_{max} = 1.924 \times 10^{-5} \, \text{Hz}$, meaning the default shortest detectable period of \textsc{GTLS} is approximately $\frac{1}{f_{max}} = 0.6$ days. This range encompasses most known transiting exoplanets; therefore, the default settings provide a practical search range even when accurate stellar parameters are unavailable.

For a typical \textit{Kepler} observation target, which has an observation length of approximately 1500 days, substituting this into Eq.~(\ref{eq:period_sampling_N}) yields $N \approx 190,000$. This means \textsc{GTLS} needs to fold its light curve 190,000 times to find the optimal solution.

\subsubsection{Duration sampling}
\label{sec:duration_sampling}

The transit duration $T_{14}$ is the time interval during which the stellar disk is occulted by the transiting object, and it directly determines the width of the transit signal. Transit duration is statistically correlated with orbital period: shorter-period planets generally produce shorter transits, whereas longer-period planets tend to produce longer transits. Therefore, the duration grid should be constructed according to the expected period--duration relation of transiting systems, so that the searched duration range remains physically consistent with real transit events and improves the match between the search template and the observed light curve.

In this work, we use $D$ to denote the number of data points in the light curve that fall within one transit duration. To avoid confusion, we use $T_{14}$ to represent the physical transit duration.

Due to the potentially relatively large inclination in the planetary orbit, which could cause the planet's transit path to graze the edge of the star, the theoretical lower limit of the duration is zero. For planets with circular orbits, the upper limit of the duration can be determined by the radii of the planet and the star, as well as the transit period, providing a more specific range.

H19 analyzed the upper limit situation and derived the following formula for the duration $T_{14}$:

\begin{equation}
	\label{eq:duration_sampling_upper_formula_raw}
	T_{14} \leq (R_{s} + R_{p})\left(\frac{4P}{\pi GM}\right)^{1/3},
\end{equation}
where $R_s$ is the radius of the star, $R_p$ is the radius of the planet.

After deriving the expression for the upper limit of the transit duration, the remaining task is to specify the values of $R_s$, $R_p$, and $P$ under suitable limiting astrophysical cases. These choices define the maximum duration to be included in the search grid. In the \textsc{GTLS} implementation, however, the light curve has already been folded over a selected trial period ($P$) when the duration grid is constructed. Therefore, for each trial period, only the adopted stellar and planetary radius limits are needed to determine the corresponding upper bound on the duration. This is the approach adopted by H19, where the relevant radius limits were chosen empirically based on astrophysical considerations and observed transiting systems.

Following H19, these boundaries should be regarded as empirical rather than as the parameters of a specific physical main-sequence system. Their purpose is to cover the observed range of known transiting exoplanets and to provide a practical period--duration envelope for the search. Since \textsc{GTLS} only requires an approximate upper relation between transit duration and orbital period, we do not impose a boundary condition corresponding to a unique physical stellar system. Instead, all quantities in Eq.~(\ref{eq:duration_sampling_upper_formula_raw}), except $T_{14}$ and $P$, are absorbed into an empirical constant ($k_{max}$). This gives

\begin{equation}
	\label{eq:duration_sampling_upper_formula_update}
		\frac{T_{14}}{P} \leq k_{max}P^{-2/3}.
\end{equation}

\begin{figure}
	\includegraphics[width=\hsize]{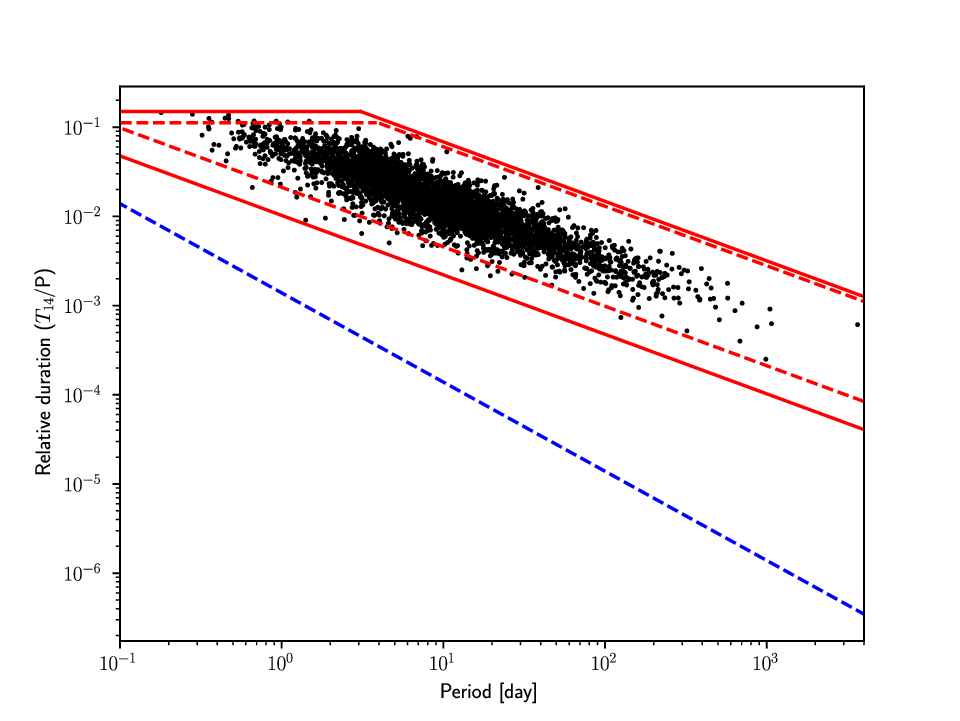}
	\caption{Transiting planets from the NASA Exoplanet Archive shown in the $T_{14}/P$ -- $P$ diagram, where  $T_{14}/P$ is the transit duration and $P$ is the orbital period. The red dashed lines indicate the boundaries of the default TLS search range, while the solid red lines mark the boundaries of the default \textsc{GTLS} search range. The blue dashed line shows the $T_{14}/P$ -- $P$ relation corresponding to a transit duration of 2 minutes, shown for reference. The data were retrieved on 2023 October 9.}
   \label{fig:duration_P}
\end{figure}

As shown in Fig.~\ref{fig:duration_P}, we performed a statistical analysis of planets in the NASA Exoplanet Archive\footnote{\url{https://exoplanetarchive.ipac.caltech.edu/}} for which both the orbital period and transit duration are available or can be derived. The figure shows the observed relationship between $T_{14}/P$ and $P$, together with the adopted boundary conditions for the \textsc{GTLS} duration search range.

Fig.~\ref{fig:duration_P} also shows that, at short orbital periods, few known exoplanets occupy the region with very small  $T_{14}/P$. This may be related to the finite time resolution and cadence limitations of current transit surveys. As a reference, the blue dashed line in Fig.~\ref{fig:duration_P} shows the relation between $T_{14}/P$ and $P$, for a transit duration of 2 minutes. In this case, a single transit would contain only about one sampling point in \textit{TESS} full-frame image data and approximately two sampling points in \textit{Kepler} short-cadence data, making such short-duration events difficult to detect reliably.

From a data-driven perspective, the lower boundary of the duration search range can be defined in the same functional form as the upper boundary by adjusting the empirical coefficient in Eq.~(\ref{eq:duration_sampling_upper_formula_update}):\begin{equation}
\label{eq:duration_sampling_lower_formula_update}
\frac{T_{14}}{P} \geq k_{min}P^{-2/3}.
\end{equation}
Although this lower boundary is empirical and does not necessarily correspond to a strict astrophysical limit, an appropriate choice of $k_{\min}$ allows the search grid to include the known transiting exoplanet population while excluding unrealistically short durations. This reduces the search space and improves computational efficiency without sacrificing sensitivity to physically relevant transit signals.

H19 adopted a similar empirical approach to extend Eq.~(\ref{eq:duration_sampling_upper_formula_raw}) and define the lower bound of the transit-duration grid. In addition, H19 imposed a maximum value of $T_{14}/P = 1.12 \times 10^{-1}$ because no known transiting exoplanets at that time had $T_{14}/P$ above this value.

Using the TLS boundary conditions and substituting the corresponding physical quantities into the duration-limit formula gives $k_{\min}=41.39\,{\rm s}^{2/3}$ and  $k_{\max}=546.94\,{\rm s}^{2/3}$. However, recently discovered transiting planets extend beyond part of the default TLS duration range. We therefore expand the default \textsc{GTLS} duration grid to better match the current exoplanet population. Based on the updated distribution of known transiting planets, we adopt $k_{\min}=20.05\,{\rm s}^{2/3}$ and $k_{\max}=620.84\,{\rm s}^{2/3}$, and increase the maximum allowed value of $T_{14}/P$ to $1.5 \times 10^{-1}$.

\subsection{Highly parallelized implementation on GPU}
\label{sec:highly_parallelized_implementation_on_gpu}

In this subsection, we discuss the key components of the search process and methods for achieving workload balance. The core transit search algorithm conducted on the GPU consists of four main steps.

The first step is to fold the transit over a wide range of trial periods using the grid defined in Sect.~\ref{sec:period_sampling}. Phase folding coherently enhances the SNR of the transit signal by averaging repeated events. The optimal orbital period is then selected from this set of trial periods.

The second step involves applying padding to the folded signal, preventing the detection algorithm from overlooking transit features located at the boundaries of the phase-folded light curve.

The third step is to compute the moving average of the light curve, which is used to estimate the local transit depth at each phase. This procedure enables appropriate scaling of the ideal transit model for direct comparison with the observed light curve.

The fourth step is to compute the transit loss $\chi^2$, which constitutes the core of the detection algorithm. Specifically, the point-by-point differences between the scaled transit model and the observed light curve are calculated and summed to obtain a $\chi^2$ value. For each trial combination of period, duration, and $T_0$, a corresponding $\chi^2$ is computed and recorded. After detrending the $\chi^2$ series, the parameter set associated with the minimum detrended $\chi_d^2$ is selected as the optimal transit signal candidate.

The first three steps of the transit search constitute a highly repetitive and non-branching computational workflow. Such operations are well suited to the single-instruction, multiple-thread (SIMD) architecture of modern graphics processing units (GPUs). However, owing to the variable execution time associated with different trial parameters, the fourth step of the search can lead to workload imbalance, as discussed in Sect.~\ref{sec:calc_chi2}.


\subsubsection{Folding the light curve}
Phase folding the light curve over a range of trial periods constitutes the first stage of the GPU-based parallelization. For a given trial period $P$, the time of each data point is divided by $P$, and the fractional part of the result is taken as the normalized orbital phase. The light curve is then sorted according to these phases to produce the folded light curve. Let $t_i$ denote the time of the $i$-th data point and $p_i$ its corresponding normalized phase. The phase folding operation can be expressed as
\begin{equation}
p_i = \frac{t_i}{P} - \lfloor \frac{t_i}{P} \rfloor.
\end{equation}

As the folding operation must be performed over a large number of trial periods and involves an identical computational workload for each trial, this step is naturally well suited to GPU parallelization.

\subsubsection{Padding the light curve and calculating the edge effects}

As the $T_0$ is generally unknown a priori, phase folding can result in a transit signal being split across the boundaries of the folded light curve, with portions appearing at both the beginning and end of the phase domain. In such cases, a direct search for contiguous transit features may lead to genuine signals being missed or their durations being incorrectly estimated.

To mitigate this edge effect and ensure consistent treatment across all trial durations, we adopt the approach used in TLS by extracting a segment from the beginning of the folded light curve and appending it to the end of the folded profile. This padding procedure ensures that transit signals are always represented as contiguous features within the search window.

\begin{figure}
	\includegraphics[width=\hsize]{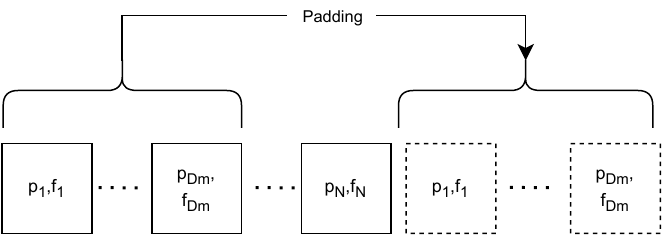}
    \caption{Padding of the folded light curve. Here, $p_i$ denotes the normalized phase of the $i$-th data point, $f_i$ is the corresponding flux value, $D_m$ is the maximum transit duration considered in the search, and $N$ is the total number of data points in the folded light curve.}
    \label{fig:patch_light_curve}
\end{figure}

Let $f_i$ denote the flux value of the $i$-th data point, $D_m$ the maximum transit duration considered in the search, and $N$ the total number of data points in the folded light curve.
This padding procedure is illustrated in Fig.~\ref{fig:patch_light_curve}.

The $\chi^2$ statistic quantifies the residuals between the folded light curve and the ideal model. When the padded light curve is compared with the ideal model, the padded segments are effectively counted twice, which leads to systematically larger $\chi^2$ values for cases with longer maximum transit durations. To correct for this bias, we subtract the contribution arising from this edge effect from the total $\chi^2$. The edge-effect term is computed using the following expression:
\begin{equation}
	\chi^2_{edge} = \sum_{i=1}^{D_m}(\frac{1 - f_{i + N - D_m}}{dy_i})^2,
\end{equation}
where $dy_i$ is the uncertainty of the $i$-th point. The detailed method for calculating $\chi^2$ will be described in Sect.~\ref{sec:calc_chi2}.
Both the padding operation and the computation of the edge-effect term are highly parallelizable and can therefore be efficiently implemented on the GPU.

\subsubsection{Calculating the moving average}

During the transit search, the transit depth is a key parameter. However, in a least-squares comparison between the ideal transit model and the observed light curve, the depth must be known in advance in order to scale the model appropriately. Unlike other transit parameters, the depth does not require explicit least-squares optimization and can instead be estimated directly from the data. For a given transit shape, the depth can be determined by computing the average flux within the transit window of the folded light curve. This provides an efficient and robust estimate of the transit depth, as expressed by the following relation:

\begin{equation}
	\frac{\delta_p}{\bar{f}_{in}} = \frac{\delta^{'}_{p}}{\bar{f}^{'}_{in}} = k,
\end{equation}
where $\delta_p$ denotes the transit depth, $\bar{f}_{in}$ is the mean flux within the observed transit window, $\delta^{'}_{p}$ is the depth of the ideal transit model, and $\bar{f}^{'}_{in}$ is the corresponding mean flux of the ideal transit model.

For an ideal transit with a fixed transit shape, the quantity $\frac{\delta^{'}_{p}}{\bar{f}^{'}_{in}}$ is a constant. To match the depth of the observed light curve, the ideal transit model must therefore be rescaled accordingly. The scaled transit depth of the ideal model is then obtained using the following relation:

\begin{equation}
	\label{eq:calc_rough_depth}
	\delta^{'}_{p} = \delta_{p} = k \cdot \bar{f}_{in}.
\end{equation}

When searching for the mid-transit time $T_0$ for a given trial period and duration, the light curve is scanned sequentially, and the sliding average is evaluated using the trial duration as the window size. As the window length varies across different trial durations, a direct implementation of the moving average would lead to a significant workload imbalance. To mitigate this, we avoid explicitly computing the moving average and instead precompute the cumulative sum of the target light curve. The moving average is then obtained by subtracting the cumulative sum at the start of the window from that at the end. This strategy ensures a uniform computational cost across trials and enables balanced workload distribution across CUDA threads.

Here, $F_{cum}(i)$ denotes the cumulative sum of the target light curve evaluated at the $i$-th data point, $F_{raw}(i)$ is the corresponding raw flux value, $F_{avg}(i)$ represents the moving average at the $i$-th point, and $D$ is the width of the moving average window. The moving average is computed using the following expression:
\begin{equation}
	\begin{array}{l}
		F_{cum}(i) = \sum_{j=1}^{i} F_{raw}(j), \\
		F_{avg}(i) = \frac{F_{cum}(i) - F_{cum}(i-D)}{D}.
	\end{array}
\end{equation}

To scale the ideal transit model, we use Eq.~(\ref{eq:calc_rough_depth}) to estimate the appropriate transit depth. The ideal transit signal is then normalized and scaled to match the corresponding mean flux level of the observed light curve.

\subsubsection{Calculating the chi-squared statistic within and outside the transit}
\label{sec:calc_chi2}

For a given trial combination of period, duration, and mid-transit time ($T_0$), the $\chi^2$ statistic can be decomposed into two components. The first term corresponds to the sum of the squared residuals between the observed light curve and the ideal transit model outside the transit window, while the second term represents the sum of the squared residuals within the transit window. For an ideal transit signal, the flux is assumed to be constant outside the transit. Since the input light curve is normalized such that the mean flux is unity, and the fractional contribution of the transit to the overall mean is negligible, the out-of-transit residuals can be approximated by subtracting the average in-transit flux from unity and squaring the result. This approximation provides an efficient and accurate estimate of the first component of the $\chi^2$ statistic.

\begin{figure}
	\includegraphics[width=\hsize]{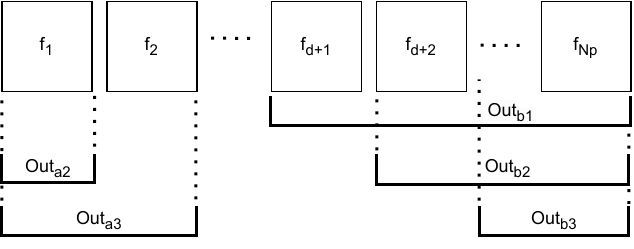}
    \caption{The composition of the $\chi^2$ value outside the transit signal. $out_a$ denotes the part before the transit signal, and $out_b$ is the part after the transit signal. As $T_0$ changes, the part before the transit signal($out_a$) is gradually increased, and the part after the transit signal($out_b$) is gradually reduced.
	$N_p$ is the number of points in the patched light curve, and $d$ is the number of points in the transit signal.}
    \label{fig:calc_chi2_first_part}
\end{figure}

The out-of-transit $\chi^2$ term can be further decomposed into two components, corresponding to the segments before and after the transit window. As illustrated in Fig.~\ref{fig:calc_chi2_first_part}, as the trial mid-transit time $T_0$ shifts from left to right, the contribution from the pre-transit segment ($\chi^2_{out,T_0,a}$) increases progressively, while the contribution from the post-transit segment ($\chi^2_{out,T_0,b}$) decreases accordingly.

For a light curve evaluated at a given trial period, duration, and  $T_0$, the out-of-transit $\chi^2$ term is computed using the following expression:
\begin{equation}
	\begin{array}{l}
	\chi^2_{out,T_0} = \chi^2_{out,T_0,a} + \chi^2_{out,T_0,b} - \chi^2_{edge}, \\
	\chi^2_{out,T_0,a} = \sum_{i=1}^{D_s - 1} (\frac{f_i - 1}{dy_i})^2, \\
	\chi^2_{out,T_0,b} = \sum_{i=D_e + 1}^{N_p} (\frac{f_i - 1}{dy_i})^2 .
	\end{array}
\end{equation}
where $D_s$ is the starting point of the transit signal, $D_e$ is the ending point of the transit signal, $N_p$ is the number of points in the patched light curve, and $dy_i$ is the uncertainty of the $i$-th point.

This formulation allows for a computationally efficient implementation since the out-of-transit $\chi^2$ values for successive trial mid-transit times ($T_0$) can be derived recursively. This only requires very low time complexity (O($n$)).

The second component of the $\chi^2$ statistic corresponds to the sum of the squared residuals between the observed light curve and the ideal transit model within the transit window. The in-transit $\chi^2$ term is computed using the following expression:
\begin{equation}
	\chi^2_{in,T_0} = \sum_{i=D_s}^{D_e} \displaystyle (\frac{f_i - f_{ideal,i}}{dy_i})^2,
\end{equation}
where $f_{ideal,i}$ denotes the flux of the ideal transit model at the $i$-th data point.

Unlike the first component of the $\chi^2$ statistic, the in-transit term cannot be computed recursively because the ideal model flux varies across the transit window. Therefore, this component must be evaluated explicitly through a direct point-by-point summation.


As discussed in Sect.~\ref{sec:duration_sampling}, the maximum transit duration scales with the trial period, while the maximum trial period scales with the total time baseline of the light curve. Consequently, the maximum transit duration is proportional to the length of the light curve and therefore to the total number of data points $n$. The computational cost of evaluating a single in-transit term, $\chi^2_{in,T0}$ , therefore scales as O($n$). Because this term must be evaluated for each possible trial mid-transit time, $T_0$, across the light curve, the overall time complexity of the in-transit $\chi^2$ calculation scales as O($n^2$).

Moreover, the trial transit durations are not uniform, which introduces a workload imbalance across GPU threads. Threads assigned to shorter durations finish their computations earlier and remain idle while waiting for threads processing longer durations to complete. As a result, evaluation of the in-transit $\chi^2$ term becomes the dominant computational bottleneck in the overall search pipeline.


\subsection{Obtaining the optimal transit signal from the chi-squared statistic}
\label{sec:detrending}

The $\chi^2$ values computed in Sect.~\ref{sec:calc_chi2} cannot be used directly to identify the optimal transit signal for two main reasons. First, an appropriate detection statistic is required to quantify the significance of a candidate signal, and such a statistic should be independent of the number of data points or the total time baseline of the input light curve. Although $\chi^2$ is the primary quantity produced by the algorithm, it does not satisfy this requirement, as it scales systematically with the length of the data set. It is therefore necessary to normalize the $\chi^2$ values in order to construct a more robust significance metric.

Following \cite{Kov_cs_2002} and H19, we define the raw signal detection efficiency (SDE) from the period-dependent signal residue. Let $\chi^2(P)$ denote the minimum $\chi^2$ obtained at a trial period $P$, and let $\chi^2_{\min}$ be the global minimum over the full period grid. The signal residue is then defined as $SR(P)=\frac{\chi^2(P)}{\chi^2_{\min}}$. We denote the mean and standard deviation of $SR(P)$ over the full period grid as $\langle SR\rangle$ and $\sigma_{SR}$, respectively. The raw SDE is then computed as
\begin{equation}
\begin{array}{l}
SDE_{\rm raw}(P) = \dfrac{SR(P)-\langle SR\rangle}{\sigma_{SR}} .
\end{array}
\end{equation}

A second limitation arises from the presence of a systematic trend in $\chi^2(P)$. As illustrated in Fig.~\ref{fig:duration_P}, the ratio $T_{14}/P$ decreases with increasing period, implying that the fraction of in-transit data points relative to the total number of points becomes progressively smaller. Consequently, a larger fraction of the ideal transit model lies  in the out-of-transit region, where the normalized model flux is unity. Because most points in the observed normalized light curve are also close to unity, the overall agreement between the model and the data improves artificially at longer periods, even when the trial period is far from the true transit period. This effect produces a systematic decrease in $\chi^2(P)$ with increasing period, which does not correspond to an increased likelihood of a genuine transit signal. Since this systematic behavior also propagates into the raw signal detection efficiency, $SDE_{raw}(P)$, it must be removed before identifying the most significant transit candidate.

Following \citet{Ofir_2014} and H19, we remove the systematic trend in $SDE_{raw}(P)$ using a moving median filter and define the resulting detrended statistic as $SDE(P)$. The optimal transit period is then taken to be the period at which $SDE(P)$ reaches its maximum value, and the corresponding duration and $T_0$ are adopted as the best-fitting transit parameters. Throughout this work, we use the peak value of $SDE(P)$ as the primary measure of transit-detection significance.

\subsection{Hierarchical search for the mid-transit time}
\label{sec:search_point_by_jumping_point_and_secondary_search}


The mid-transit time, $T_0$, is treated as an independent search parameter together with the orbital period and transit duration. Unlike the period and duration, which are searched over predefined finite grids, $T_0$ can in principle occur at any time within the light curve. To reduce the computational burden, we restrict $T_0$ to the discrete sampling of the light curve and assume that the trial mid-transit time coincides with one of the observed time points. For a typical \textit{Kepler} long-cadence light curve, this still implies that nearly $10^5$ possible $T_0$ values must be evaluated for each trial combination of period and duration. An exhaustive search over all possible $T_0$ values therefore represents a major contribution to the overall computational cost.



To improve efficiency, TLS evaluates only a subset of points when searching over $T_0$. This is done by introducing a step-size parameter,  $\Delta t_0$, which controls the spacing between successive trial mid-transit times. Equivalently,  $\Delta t_0$ down-samples the $T_0$ search grid. The number of skipped data points is given by
\begin{equation}
N_{\rm skip} = n_{\rm IT} \cdot \Delta t_0 ,
\end{equation}
where $n_{IT}$ is the number of data points spanning the trial transit duration, and $N_{skip}$ is the number of data points skipped between two consecutive trial values of $T_0$.

The value of $N_{skip}$ depends only on the prescribed $\Delta t_0$ and the tested transit duration and is independent of the trial period. Thus, for a fixed period, $N_{skip}$ varies with the trial duration. For example, if $n_{IT}$=2000 and $\Delta t_0$=0.05, then $N_{skip}=100$ indicates that 99 data points are skipped between two successive trial mid-transit times.


\textsc{GTLS} adopts this skipping strategy and extends it into a two-stage hierarchical search. In the first stage, \textsc{GTLS} performs a coarse search over $T_0$ using relatively large skip intervals across the full grid of trial periods and durations. This stage rapidly identifies promising candidate combinations and computes a preliminary goodness-of-fit statistic, such as the least-squares residual, for each case. In the second stage, \textsc{GTLS} performs a local refinement around the most promising candidates without skipping data points. During this refinement step, detection statistics such as the SNR and SDE are evaluated to determine the optimal combination of period, duration, and $T_0$. This hierarchical strategy preserves accurate transit alignment in the phase-folded light curve while substantially reducing the computational cost of the $T_0$ search.

As discussed in Sect.~\ref{sec:period_sampling}, for a typical \textit{Kepler} target, \textsc{GTLS} first performs a coarse search over approximately 190,000 folded light curves. It then applies the refinement search to roughly 200 candidate folded light curves selected according to $SDE(P)$. Because the second-stage search is restricted to a small subset of promising trial periods, its additional computational cost is negligible. Nevertheless, this refinement step significantly improves the accuracy of the recovered transit parameters.



\section{Benchmark Performance}
\label{sec:sec_performance}

In this section, we evaluate the performance of \textsc{GTLS} using both synthetic light curves and real \textit{Kepler} light curves containing Kepler Objects of Interest (KOIs). We compare \textsc{GTLS} with TLS and BLS in terms of detection sensitivity, recovery of transit parameters, and computational efficiency.

As described in Sect.~\ref{sec:period_sampling}, the default period grids adopted by \textsc{GTLS} and TLS span a sufficiently broad range to search for unknown planetary systems. To ensure an unbiased comparison under realistic blind-search conditions, we do not use prior information on stellar or planetary parameters when constructing the search grids. Instead, all methods are evaluated over the same period search space.

As discussed in Sect.~\ref{sec:duration_sampling}, \textsc{GTLS} explores a wider range of transit durations than TLS. For BLS, we adopt the duration search range used by \citet{Kunimoto_2023}, with commonly used parameters
\footnote{\texttt{q\textunderscore min} = 2e-4, \texttt{qmax} = 0.15, \texttt{dlogq} = 0.1, and \texttt{noverlap} = 3.}.

\subsection{Detection Sensitivity}
\label{sec:sec_performance_detection_sensitivity}

Transit-detection performance can be assessed from two complementary perspectives. The first is the ability to distinguish light curves containing transit signals from those without detectable transit signals, which is commonly quantified using detection statistics such as the SNR and SDE. The second is the accuracy with which the fundamental transit parameters are recovered, including the orbital period, transit duration, transit depth, and $T_0$. In this subsection, we first assess the detection sensitivity of \textsc{GTLS} and TLS using synthetic white-noise light curves.

Following the tests presented in H19, we generate synthetic light curves containing Earth-like transit signals using \texttt{batman} \citep{Kreidberg_2015}. Gaussian white noise with an amplitude of 110 ppm is added to mimic the typical 30-min noise level of \textit{Kepler} observations for a $K = 12$ mag solar-type star \citep{Gilliland_2015}. To match \textit{Kepler} long-cadence observations, each simulated light curve spans four years and is sampled every 30 min.

We design two complementary tests to evaluate the performance of \textsc{GTLS}. The first uses synthetic light curves with injected white noise to test the idealized detection sensitivity of the algorithm under controlled conditions. The second uses real \textit{Kepler} light curves containing KOIs to evaluate recovery performance under realistic observational noise and systematics. Together, these tests quantify both the intrinsic algorithmic performance of \textsc{GTLS} and its practical robustness when applied to real photometric data.

We evaluate the signal-recovery capability of \textsc{GTLS} and TLS using SDE and SNR. The two methods produce statistically similar SDE and SNR distributions, with comparable means and standard deviations. This indicates that \textsc{GTLS} preserves the detection sensitivity of TLS for  identifying  transit signals in white-noise light curves.

\begin{table}
	\centering
    \caption{The precision and recall of \textsc{GTLS} and TLS using the common detection thresholds of SDE = 7 and SNR = 5.}
    \label{table:PrecisionAndRecall}
	\begin{tabular}{lcc} 
	 \hline\hline
	 Parameter & \textsc{GTLS} & TLS \\
	 \hline
	 Precision & 9.3\% & 9.4\% \\ 
	 Recall & 79.4\% & 81.1\% \\
	 \hline
	\end{tabular}
\end{table}

In practice, fixed thresholds in SDE and SNR are commonly adopted to define candidate KOIs. The resulting precision and recall values 
for \textsc{GTLS} and TLS  are summarized in Table~\ref{table:PrecisionAndRecall}. These results show that \textsc{GTLS} and TLS achieve comparable detection performance, with statistically similar precision and recall. The precision and recall are defined as:
\begin{equation}
	\begin{array}{l}
		\mathrm{Precision} = \frac{TP}{TP + FP}, \\[4pt]
		\mathrm{Recall} = \frac{TP}{TP + FN}.
	\end{array}
\end{equation}
Here, $TP$ is the number of true positives, $FP$ is the number of false positives, and $FN$ is the number of false negatives. Positive cases correspond to light curves containing KOIs with known orbital periods shorter than 14 days, while negative cases correspond to  light curves without such KOIs.

Both \textsc{GTLS} and TLS yield relatively low precision values: 9.3 percent and 
9.4 percent, respectively. These low precision values should not be interpreted as poor final validation performance. Like TLS and BLS, \textsc{GTLS} is intended as a high-recall first-stage search algorithm. Candidates identified at this stage require subsequent vetting, false-positive rejection, and astrophysical validation. The precision values reported here reflect intentionally permissive first-stage detection thresholds chosen to maintain high completeness. They should therefore not be interpreted as final planet-candidate reliability values, since centroid vetting,  odd-even tests, secondary-eclipse searches, astrophysical false-positive rejection, and other validation steps are not included in this benchmark.

\subsection{ROC Comparison of \textsc{GTLS}, TLS, and BLS}
\label{sec:sec_performance_ROC}

\begin{figure}
	\centering
	\includegraphics[width=\hsize]{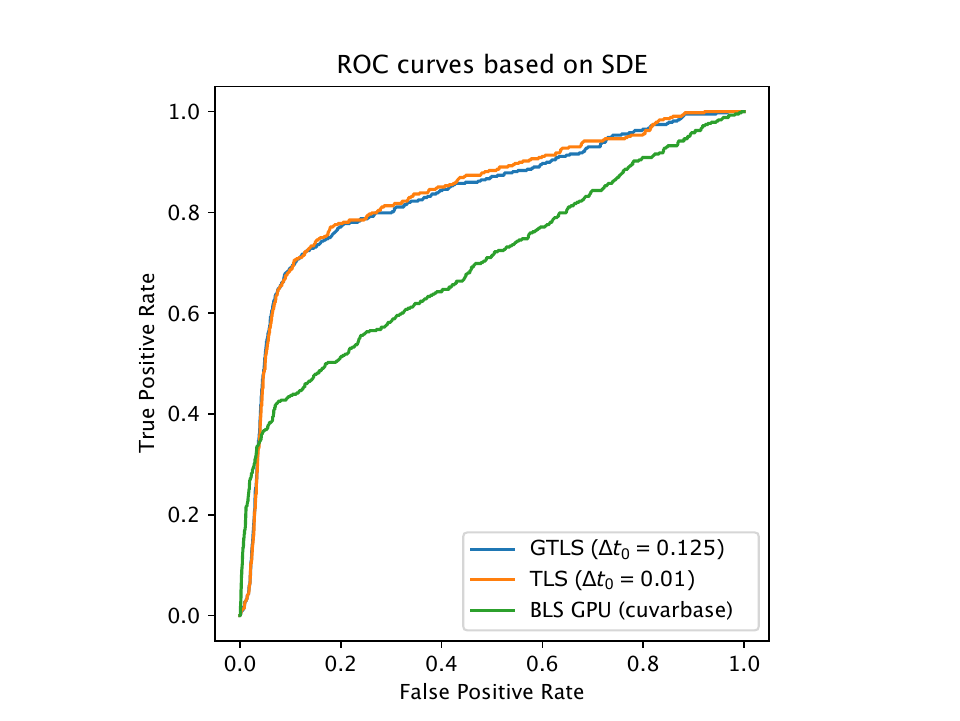}
	\caption{
	The receiver operating characteristic (ROC) curves for \textsc{GTLS}, TLS, and BLS. 
	The closer the ROC curve is to the upper-left corner, the better the diagnostic 
	ability of the test. See the text for more details.
	}
	\label{fig:ROC_curve}
\end{figure}

To further assess the consistency between \textsc{GTLS} and TLS, we use SDE as the detection statistic and construct receiver operating characteristic (ROC) curves for \textsc{GTLS}, TLS, and BLS, as shown in Fig.~\ref{fig:ROC_curve}. The ROC curve 
shows the true positive rate (TPR) as a function of the false positive rate (FPR) for different decision thresholds:
\begin{equation}
	\begin{array}{l}
		\mathrm{TPR} = \frac{TP}{TP + FN}, \\[4pt]
		\mathrm{FPR} = \frac{FP}{FP + TN}.
	\end{array}
\end{equation}
The true positive rate is equivalent to the recall, while the false positive rate measures the fraction of negative samples that are incorrectly classified as positive. The ROC curve, therefore, provides a threshold-independent measure of the diagnostic performance of each detection method.

As shown in Fig.~\ref{fig:ROC_curve}, the ROC curves of \textsc{GTLS} and TLS are nearly identical, indicating that the two methods have statistically indistinguishable discriminative performance. Over a broad range of false positive rates, both \textsc{GTLS} and TLS outperform BLS.  At very low false positive rates ($\mathrm{FPR} < 0.1$), BLS achieves slightly higher TPR values, suggesting a 
marginally stronger discriminative capability in this limited regime. However, all three methods have relatively low TPRs at such low FPR levels.

This behavior reflects the role of \textsc{GTLS}, TLS, and BLS as first-stage  screening tools in transit detection pipelines. At this stage, the primary goal is not to minimize the false positive rate at the expense of completeness, but to maintain a high true positive rate and avoid missing genuine transit signals. In practice, this motivates the use of relatively permissive detection thresholds, which increase 
sensitivity while also producing more false positives. Under this operational regime, \textsc{GTLS} and TLS show a clear advantage over BLS in recovering potential transit candidates.

\subsection{Recovery of KOI Signals and Transit Parameters}
\label{sec:sec_performance_accuracy_KOI}

We further evaluate the ability of \textsc{GTLS}, TLS, and BLS to recover transit or transit-like signals from real \textit{Kepler} light curves. This experiment tests not only whether a signal can be detected but also how accurately the recovered transit parameters match reference KOI values.

An increasing number of machine-learning approaches \citep{Shallue_2018,Ansdell_2018,Rao_2021} have been developed to automatically analyze 
and validate transit signals. These methods typically require pre-determined scalar transit parameters, including orbital period, $T_0$, transit duration, and transit depth. The accuracy of these parameters directly affects the quality of light-curve preprocessing and, consequently, the performance of downstream classification and validation.

Owing to the diversity and heterogeneity of photometric time series, \textsc{GTLS}---like BLS and TLS---does not include an intrinsic preprocessing module. Nevertheless, some preprocessing is required to remove long-term trends and to normalize 
the light curves to a consistent scale. Since preprocessing is not the primary focus of this work, and \textsc{GTLS} is designed to maximize computational efficiency, we adopt a simple and broadly applicable preprocessing scheme.

The preprocessing consists of two steps. First, using the known stellar parameters, we estimate the expected transit duration and apply a detrending window equal to three times this duration. The trend is removed using the biweight filter  implemented in \texttt{wotan} \citep{Hippke_2019b}. Second, light curves from different \textit{Kepler} quarters for the same target are stitched together and concatenated into a single time series.

We note that light-curve preprocessing is inherently non-trivial and can significantly influence transit detection performance. In this work, the adopted preprocessing scheme is intended to remove dominant long-term trends in a computationally efficient manner, rather than to provide optimized  detrending for individual targets.

For our evaluation, we selected the first 2,815 targets from the KOI catalog in chronological order. For each target, we concatenate the long-cadence light curves from different \textit{Kepler} quarters and apply the preprocessing procedure described above. We then apply \textsc{GTLS}, TLS, and the BLS implementation from \texttt{cuvarbase} to the preprocessed light curves and compare their performance.

By construction, the KOI sample contains light curves with at least one detected transit or transit-like event. Although some of these signals have subsequently been identified as false positives, distinguishing  astrophysical planets from false positives  is beyond the scope of these algorithms, which are designed as preliminary screening tools. Therefore, this comparison focuses on the ability of the three methods to recover transit or transit-like signals, as quantified by the accuracy of the recovered transit parameters.

\subsubsection{Average Relative Error of Recovered Parameters}
\label{sec:sec_performance_ARE}

We compute the average relative error (ARE) of the parameters recovered by \textsc{GTLS}, TLS, and BLS as follows:
\begin{equation}
	\label{eq:ARE}
	\begin{array}{l}
		ARE_{P} = \frac{\left|P_{r} - P_{e}\right|}{\max(P_{r},P_{e})}, \\[6pt]
		ARE_{D} = \frac{\left|D_{r} - D_{e}\right|}{\max(D_{r},D_{e})}, \\[6pt]
		ARE_{Dp} = \frac{\left|Dp_{r} - Dp_{e}\right|}{\max(Dp_{r},Dp_{e})}, \\[6pt]
		ARE_{T0} = 
		\frac{\left|T0_{r} - \left(T0_{e} + nP_{e}\right)\right|}{P_{e}} .
	\end{array}
\end{equation}
Here, the subscript $r$ denotes the recovered value, and the subscript $e$ denotes the reference value obtained from ExoFOP. The parameters $P$, $D$, $Dp$, and $T0$ represent the orbital period, transit duration, transit depth, and mid-transit time, respectively. When multiple KOIs are associated with the same target, we select the KOI whose reported period is closest to the recovered period for comparison.

For the period error, we normalize by $\max(P_{r},P_{e})$ to ensure that $ARE_{P} < 1$. The same normalization is applied to  $ARE_{D}$ and $ARE_{Dp}$. Since the reference $T0_{e}$ may lie outside the time baseline of  the observed light curve, we shift it by an integer number of periods, $nP_{e}$, so that $T0_{e} + nP_{e}$ falls within the observed time range and is as close as possible to the recovered $T0_{r}$.

The BLS implementation in \texttt{cuvarbase} does not natively provide a transit-depth  estimate. We therefore compute the depth from the recovered transit parameters:
\begin{equation}
	\label{eq:depth}
	\delta = 1 - \bar{F}_{IT},
\end{equation}
where $F_{IT}$ denotes the in-transit flux value and $\bar{F}_{IT}$ is their 
mean. Owing to limitations of the BLS algorithm, the recovered transit depth may 
occasionally be negative. In such cases, we set the depth to zero, which by definition 
gives $ARE_{Dp} = 1$.

\begin{figure}
	\centering
	\includegraphics[width=\hsize]{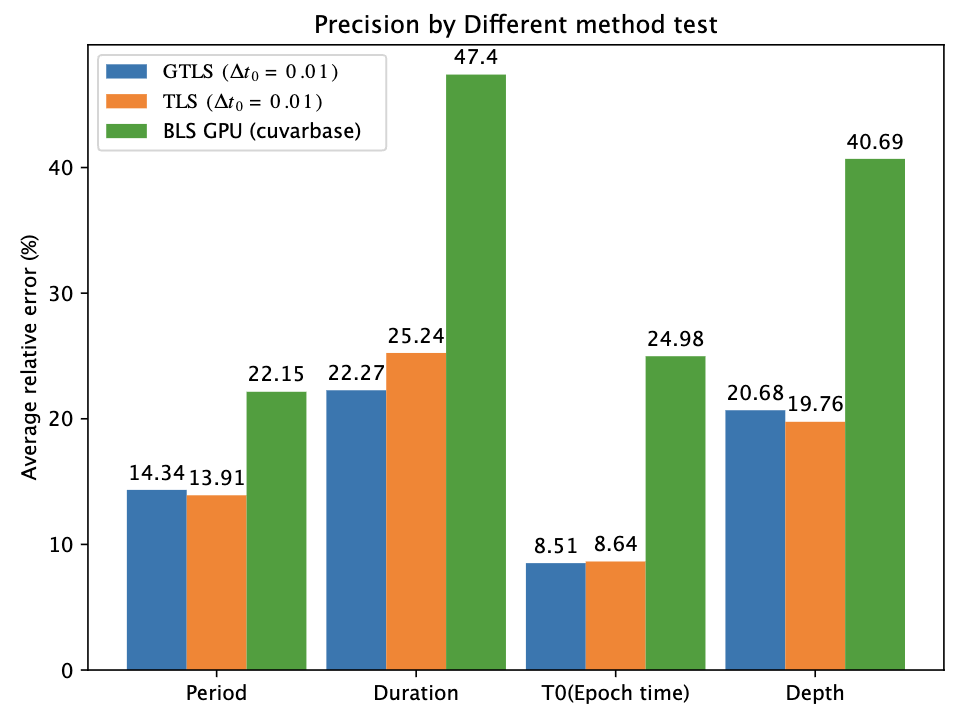}
    \caption{
    Average relative errors of the transit parameters recovered by \textsc{GTLS}, TLS, 
    and BLS. Lower values indicate more accurate parameter recovery.
    See the text for more details.
    }
	\label{fig:AverageRelativeError}
\end{figure}

Fig.~\ref{fig:AverageRelativeError} shows the AREs of the recovered transit parameters for \textsc{GTLS}, TLS, and BLS. Both \textsc{GTLS} and TLS outperform BLS across all parameters; although \textsc{GTLS} shows slightly lower accuracy than TLS in recovering orbital periods for KOI light curves, it achieves better  recovery of the transit duration,  $T_0$, and transit depth.

\subsection{Computational Efficiency}
\label{sec:sec_performance_speed}

We compare the computational performance of \textsc{GTLS}, TLS, and the GPU-based BLS implementation from \texttt{cuvarbase} using synthetic transit light curves. The light curves are generated with \texttt{batman} \citep{Kreidberg_2015}, sampled at \textit{Kepler's} long-cadence resolution (30 min) and constructed with different total time baselines. The tested baselines include a \textit{Kepler-like} duration of approximately 1500 d and an extended 3000-d case.

For each test, the runtime is computed by subtracting the time spent on external preprocessing from the total execution time of each algorithm. TLS is run on an AMD Ryzen 9 7950X CPU, while \textsc{GTLS} and BLS are run on an NVIDIA GeForce RTX 4090 GPU. These platforms  represent state-of-the-art consumer-grade hardware available at the time of writing.

\begin{figure}
	\centering
	\includegraphics[width=\hsize]{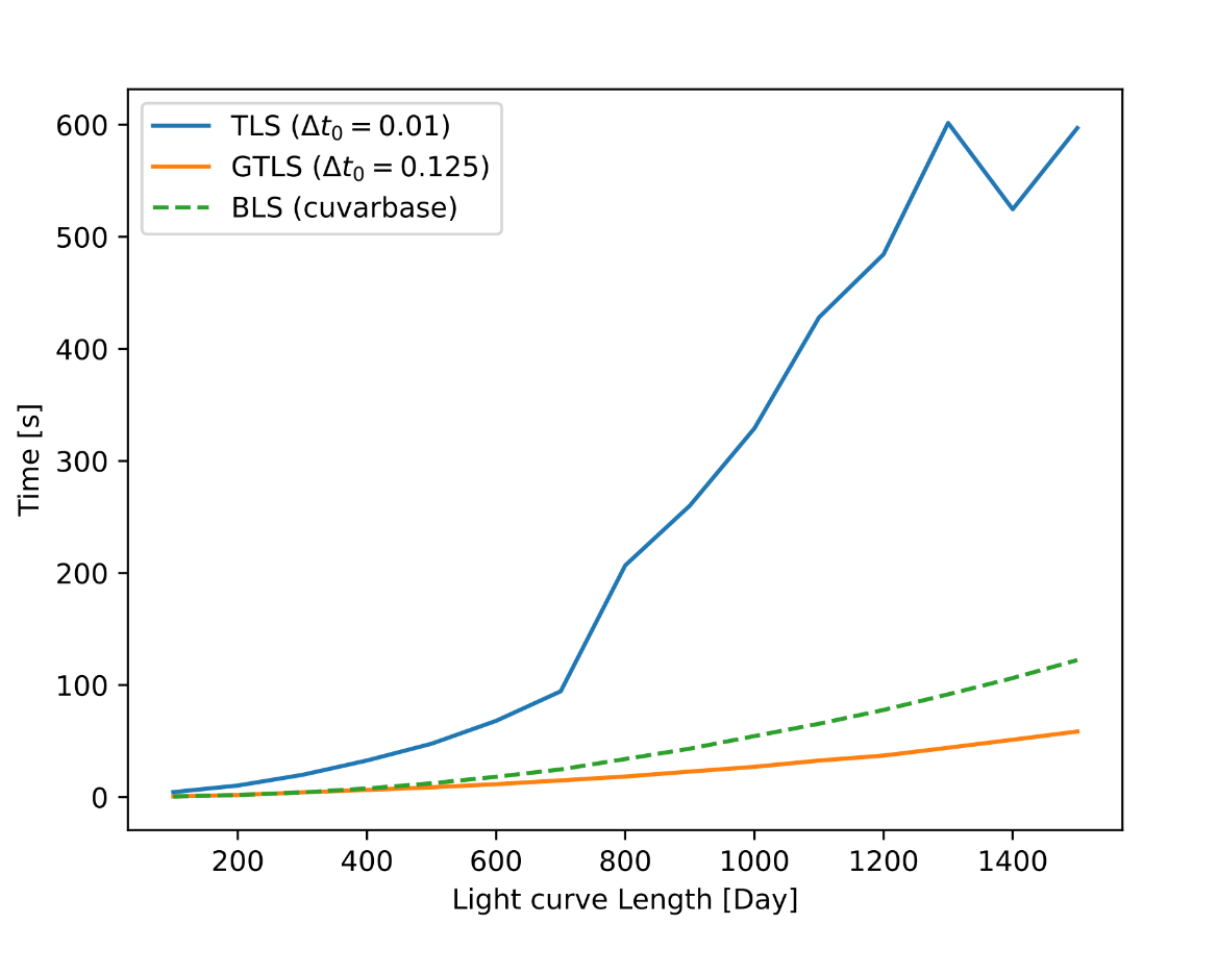}
    \caption{
Runtime comparison among \textsc{GTLS}, TLS, and the GPU-based BLS implementation in \texttt{cuvarbase}. \textsc{GTLS} is significantly faster than both comparison methods over the tested light-curve baselines.
    }
	\label{fig:speedCompare}
\end{figure}

For a \textit{Kepler-like} light curve spanning 1500 d, \textsc{GTLS} completes the search in 33.3 s, compared with 522.0 s for TLS and 121.1 s for GPU-BLS. This corresponds to speed-up factors of 15.7 relative to TLS and  3.6 relative to BLS (121.1 s) (Fig.~\ref{fig:speedCompare}).

This demonstrates the substantial computational advantage of \textsc{GTLS} over both CPU-based TLS and GPU-based BLS. The performance gain becomes even more pronounced for longer time series. For a synthetic 3000-d  light curve, \textsc{GTLS} requires approximately 138 s using a single GPU, whereas TLS requires 3289 s. In addition,  \textsc{GTLS} supports multi-GPU parallelization: on a dual RTX 4090 configuration, the runtime for the 3000-d light curve is further reduced to 79 s, corresponding to  an additional speed-up of approximately 1.75 relative to the single-GPU case.

\section{Discussion}

In this paper, the term ``TLS-like sensitivity'' refers to the fact that \textsc{GTLS} produces statistically similar SDE and SNR distributions, ROC behavior, and KOI recovery performance to the reference TLS implementation under the benchmark settings adopted in this paper. It does not imply bit-level agreement with TLS outputs.

\subsection{Remaining computational bottlenecks}

As discussed in Sect.~\ref{sec:calc_chi2}, the dominant computational cost in TLS-style searches arises from the evaluation of the $\chi^2$ statistic. To quantify this cost in \textsc{GTLS}, we analyze the runtime breakdown  for a \textit{Kepler} long-cadence light curve spanning 3000 d. The results are summarized in Table~\ref{tab:runtime_breakdown}. 

We find that approximately 40.4 percent of the total execution time is spent computing the in-transit $\chi^2$ component, making this step  the principal performance bottleneck. This bottleneck arises because the in-transit $\chi^2$ term must be evaluated through direct point-by-point summation, and because different trial transit durations lead to uneven workloads across GPU threads. Further acceleration of TLS-style algorithms will therefore require improved strategies for mitigating this workload imbalance. 

\begin{table}
	\centering
	\caption{Runtime breakdown of \textsc{GTLS} for a 3000-d \textit{Kepler}-like long-cadence light curve.}
	\label{tab:runtime_breakdown}
	\begin{tabular}{lc}
		\hline\hline
		Computational step & Runtime fraction \\
		\hline
		In-transit $\chi^2$ calculation & 40.4\% \\
		Cumulative sum of the folded light curve & 28.9\% \\
		Out-of-transit $\chi^2$ calculation & 11.5\% \\
		Other operations & 19.2\% \\
		\hline
	\end{tabular}
\end{table}

\subsection{Choice of Delta t0}
\label{sec:t0_choice}

As discussed in Sect.~\ref{sec:search_point_by_jumping_point_and_secondary_search}, $n_{IT} \cdot \Delta t_0$ defines the number of data points skipped between successive trial values of $T_0$. The choice of $\Delta t_0$ therefore represents a trade-off between computational efficiency and detection sensitivity. Smaller values of $\Delta t_0$ sample $T_0$ more densely and improve transit alignment, but they increase the computational workload. Larger values reduce the number of trial epochs, but may miss or misalign transit signals, thereby reducing detection sensitivity (H19).

This trade-off is particularly important for \textsc{GTLS}.  To quantify its effect, we tested the runtime sensitivity of both \textsc{GTLS} and TLS  using synthetic light curves with a duration of 1500 days and a cadence of 30 minutes. Reducing $\Delta t_0$ from 0.1 to 0.01 increases the runtime of \textsc{GTLS} by approximately 37 percent, whereas the corresponding increase for TLS is only about 8 percent. This indicates that the computational cost of \textsc{GTLS} is more sensitive to the adopted value of $\Delta t_0$.

The reason is that the dominant computational bottleneck in \textsc{GTLS} is the in-transit  $\chi^2$ calculation, whose cost scales directly with the number of evaluated $T_0$ values. In contrast, the runtime of CPU-based TLS is distributed across several computational components, making the effect of  $\Delta t_0$ less pronounced. Therefore, when applying \textsc{GTLS}, users should choose $\Delta t_0$ according to the desired balance between detection sensitivity and computational efficiency.

\subsection{Limitations}
\label{sec:limitations}

Although \textsc{GTLS} achieves significant computational speed-ups while preserving  TLS-like detection sensitivity, several limitations should be considered  when applying the method to large-scale transit surveys.

\textbf{Hardware and software dependencies:} The current implementation of \textsc{GTLS} relies on \texttt{CuPy} and the CUDA framework for GPU acceleration. As a result, the full acceleration capability is currently limited to  NVIDIA GPUs. Porting \textsc{GTLS}  to other  hardware architectures, such as AMD GPUs or Apple Silicon, would require substantial adaptation of the computational backend.

\textbf{Transit Timing Variations (TTVs):} Like standard TLS and BLS, \textsc{GTLS} assumes strictly periodic transits. For  systems with strong TTVs, mutual gravitational interactions can shift individual transit times away from a strictly periodic ephemeris. This can cause misalignment during phase folding,   degrade the in-transit $\chi^2$ evaluation, and suppress the resulting SDE.

\textbf{Long-period and sparse-transit signals:} Detecting long-period exoplanets with only two observable transits remains intrinsically difficult. In this regime, the number of in-transit data points is small, and the folded transit signal is more vulnerable to local noise fluctuations and isolated outliers. This can increase the probability of missed detections or poorly constrained parameters.

\textbf{GPU memory usage:} To achieve high speed, \textsc{GTLS} evaluates large  grids of periods, durations, and trial mid-transit times in parallel. Consequently, GPU memory usage increases  with the length of the input light curve and the resolution of the search grids. Very long time series or high-cadence observations  may therefore require data chunking, grid downsampling, or other memory-management strategies to remain within the limits of consumer-grade GPUs.

\section{Conclusions}

Efficient and accurate transit detection is increasingly important for modern large-scale photometric surveys. In this work, we present \textsc{GTLS}, a GPU-accelerated implementation of the Transit Least Squares algorithm designed to accelerate TLS-style periodic transit searches while preserving sensitivity to transit-shaped signals.

Our benchmarks show that \textsc{GTLS} provides a substantial improvement in computational performance relative to the original CPU-based TLS implementation. It also outperforms the GPU-based BLS implementation in \texttt{cuvarbase} in runtime under the tested conditions. For a \textit{Kepler-like} 1500-d light curve, \textsc{GTLS} completes the search in 33.3 s, corresponding to a speed-up of 15.7 relative to TLS and 3.6 relative to GPU-BLS. For a 3000-d light curve, \textsc{GTLS} requires approximately 138 s on a single RTX 4090 GPUs, and this runtime is reduced to 79 s when two RTX 4090 GPUs are used.

In terms of detection performance, \textsc{GTLS} produces SDE and SNR distributions, ROC behavior, and KOI recovery performance that are statistically consistent with TLS under the adopted benchmark settings. Both \textsc{GTLS} and TLS also show better recovery performance than BLS across the tested data sets, particularly for transit-shaped signals where the TLS model provides a better match than a box-shaped template.

Overall, \textsc{GTLS} provides an efficient and practical framework for large-scale transit searches, offering a favorable balance between computational speed and detection sensitivity. Its GPU-based implementation makes TLS-style searches more feasible for large photometric data sets from \textit{Kepler}, TESS, PLATO, ET, and future time-domain surveys.

\begin{acknowledgements}

Funding for this study is provided by the Strategic Priority Program on Space Science of the Chinese Academy of Sciences (XDA15020600) and China's Space Origins Exploration Program (GJ11030405).

This paper includes data collected by the \textit{Kepler} mission and obtained from the MAST data archive at the Space Telescope Science Institute (STScI). Funding for the \textit{Kepler} mission is provided by the NASA Science Mission Directorate. STScI is operated by the Association of Universities for Research in Astronomy, Inc., under NASA contract NAS 5--26555.

This research has made use of the Exoplanet Follow-up Observation Program (ExoFOP; DOI: 10.26134/ExoFOP5) website,
which is operated by the California Institute of Technology, under contract with the National Aeronautics and Space Administration under the Exoplanet Exploration Program.

This research has made use of the NASA Exoplanet Archive, which is operated by the California Institute of Technology,
under contract with the National Aeronautics and Space Administration under the Exoplanet Exploration Program.

\end{acknowledgements}

\section*{Data availability}

The \textsc{GTLS} source code, example scripts, and instructions for reproducing the benchmark tests are available at the \textsc{GTLS} project repository: \url{https://github.com/Farthing-0/GTLS}. The \textit{Kepler} light curves used in this work are publicly available from MAST, and KOI parameters were obtained from ExoFOP and the NASA Exoplanet Archive.



\bibliographystyle{aa}
\bibliography{gtls}








\end{document}